\documentclass{iau}
\usepackage{graphicx}

\title[Precise Near-Infrared Radial Velocities] 
{Precise Near-Infrared Radial Velocities}

\author[P. Plavchan et al.]   
{Peter Plavchan$^1$, Peter Gao$^2$, Jonathan Gagne$^3$, Elise Furlan$^4$, Carolyn Brinkworth$^5$, Michael Bottom$^2$, Angelle Tanner$^6$, Guillem Anglada-Escude$^7$, Russel White$^8$, Cassy Davison$^8$, Sean Mills$^9$, Chas Beichman$^4$, John Asher Johnson$^{10}$, David Ciardi$^4$, Kent Wallace$^{11}$, Bertrand Mennesson$^{11}$, Gautam Vasisht$^{11}$, Lisa Prato$^{12}$, Stephen Kane$^{13}$, Sam Crawford$^{11}$, Tim Crawford$^{11}$, Keeyoon Sung$^{11}$, Brian Drouin$^{11}$, Sean Lin$^{11}$, Stephanie Leifer$^{11}$, Joe Catanzarite$^{14}$, Todd Henry$^{15}$, Kaspar von Braun$^{12}$, Bernie Walp$^{16}$, Claire Geneser$^1$, Nick Ogden$^1$, Andrew Stufflebeam$^1$, Garrett Pohl$^1$  \and Joe Regan$^1$}

\affiliation{$^1$Missouri State University\\email: {\tt plavchan@missouristate.edu}\\
$^2$Caltech, $^3$University of Montreal, $^4$NASA Exoplanet Science Institute, $^5$National Center for Atmospheric Research/ University Corporation for Atmospheric Research, $^6$Mississippi State University,$^7$University College London,$^8$Georgia State University,$^9$University of Chicago,$^{10}$Harvard,$^{11}$NASA Jet Propulsion Laboratory,$^{12}$Lowell Observatory,$^{13}$San Francisco State University,$^{14}$SETI Institute,$^{15}$Georgia State University,$^{16}$NASA Ames}

\pubyear{2015}
\volume{314}  
\pagerange{119--126}
\jname{Young Stars \& Planets Near the Sun}
\editors{J. H. Kastner, B. Stelzer, \& S. A. Metchev, eds.}
\begin{document}

\maketitle

\begin{abstract}
We present the results of two 2.3 $\mu$m near-infrared (NIR) radial velocity (RV) surveys to detect exoplanets around 36 nearby and young M dwarfs. We use the CSHELL spectrograph ($R\sim$46,000) at the NASA InfraRed Telescope Facility (IRTF), combined with an isotopic methane absorption gas cell for common optical path relative wavelength calibration. We have developed a sophisticated RV forward modeling code that accounts for fringing and other instrumental artifacts present in the spectra. With a spectral grasp of only 5 nm, we are able to reach long-term radial velocity dispersions of $\sim$20-30 m/s on our survey targets.  \\
\\
keywords: instrumentation: spectrographs, techniques: radial velocities
\end{abstract}

\vspace*{-1.3 cm}
\section{Introduction}

We are carrying out an exoplanet search to exploit the ``M dwarf opportunity.''  Approximately 70\% of main sequence stars are M dwarfs, spanning a factor of 5 in mass/radius and $\sim$10$^3$ in luminosity.  Due to their smaller masses and radii and cooler temperatures compared to FGK stars, transiting exoplanets produced deeper eclipses of M dwarfs, exoplanet habitable zones are closer, and exoplanet radial velocity amplitude reflex motions are larger.  More recently, a new ``M dwarf opportunity'' for finding exoplanets has been identified with the Kepler mission -- not only are planets smaller than 4 $R_\oplus$ more common than Jovian planets, they are predominantly found around lower mass stars (Howard et al. 2012, Dressing \& Charbonneau 2013).

There are several factors spoiling the ``M dwarf opportunity.''  First, M dwarfs are red and faint, with $V-K>3.5$ mag and only 4 M4 or later dwarfs with $V<12$ mag.  Additionally, small planetary RV signals are comparable in amplitude and time-scale to numerous periodic perturbing stellar jitter signals such as rotational spot modulation \& magnetic activity cycles (Vanderberg et al. 2015, Robertson et al. 2015).  One solution to both of these challenges is to observe in the NIR, where M dwarfs put out much of their bolometric luminosity, and the amplitude of stellar jitter is diminished (Reiners et al. 2010, Plavchan et al. 2013a,b,2015, Anglada-Escude et al. 2012).

\vspace*{-0.5 cm}
\section{Results and Conclusions} 

We have obtained our goal precision of $\sim$30-60 m/s in our young M dwarf survey from 2010-2012, and $\sim$20-30 m/s in 2014 at higher SNR for nearby M dwarfs not previously surveyed by visible RV teams.  We present a sample RV curve and our survey results in Figure 1.   We are in the process of publishing our detailed survey results.  

Looking towards the future, we have made the gas cells for iSHELL (R$\sim$75,000, spectral grasp of 250 nm), which will replace CSHELL in 2016.  We expect to get to $<$ 5 m/s precision with iSHELL due to the increased spectral grasp alone (Figure 2).   The increased efficiency, resolution, no fringing, and smaller number of bad pixels will provide further improvements.

\begin{figure}
\vspace*{-0.3 cm}
\begin{center}
\includegraphics[width=2.2in]{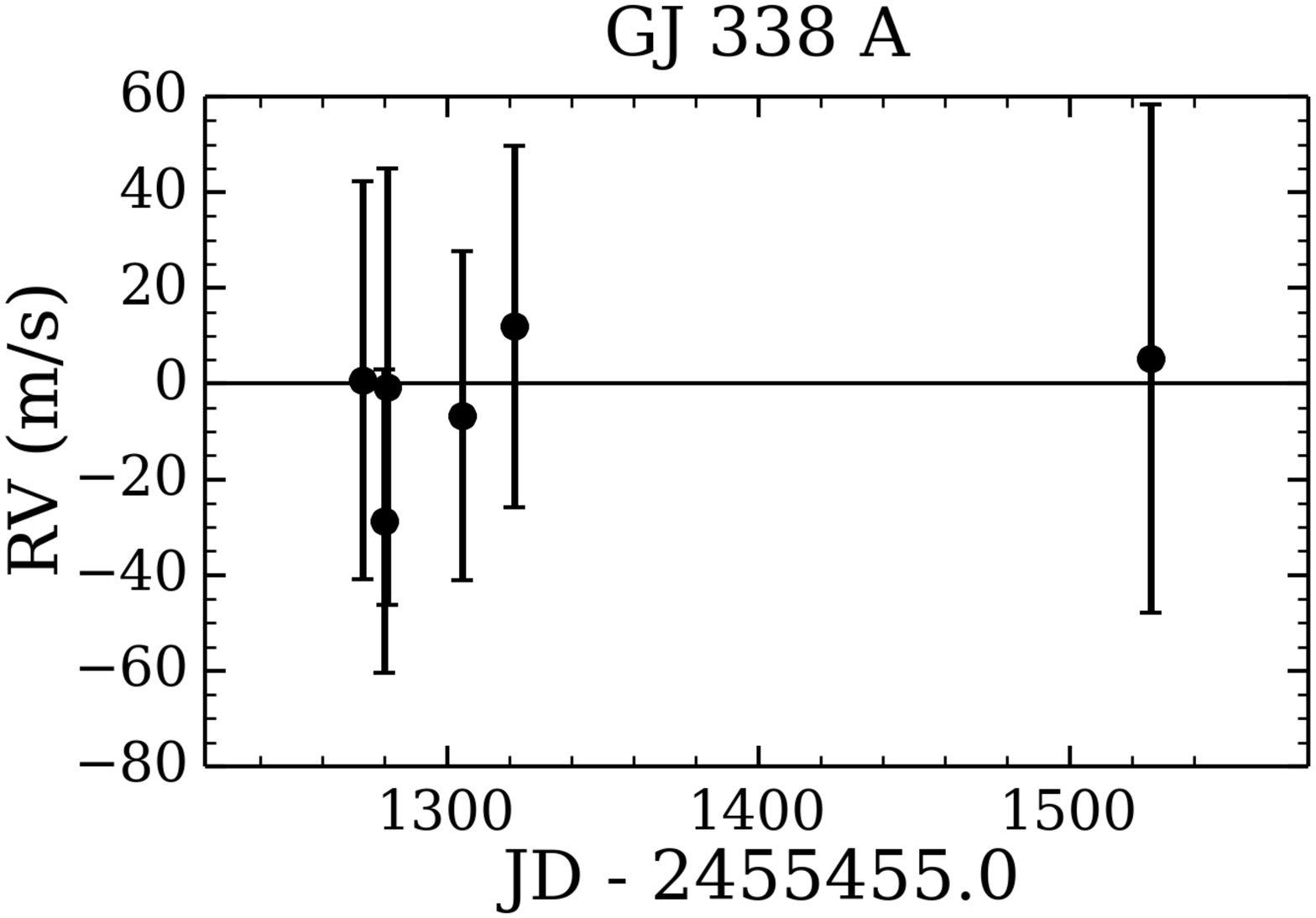} 
\includegraphics[width=2.5in]{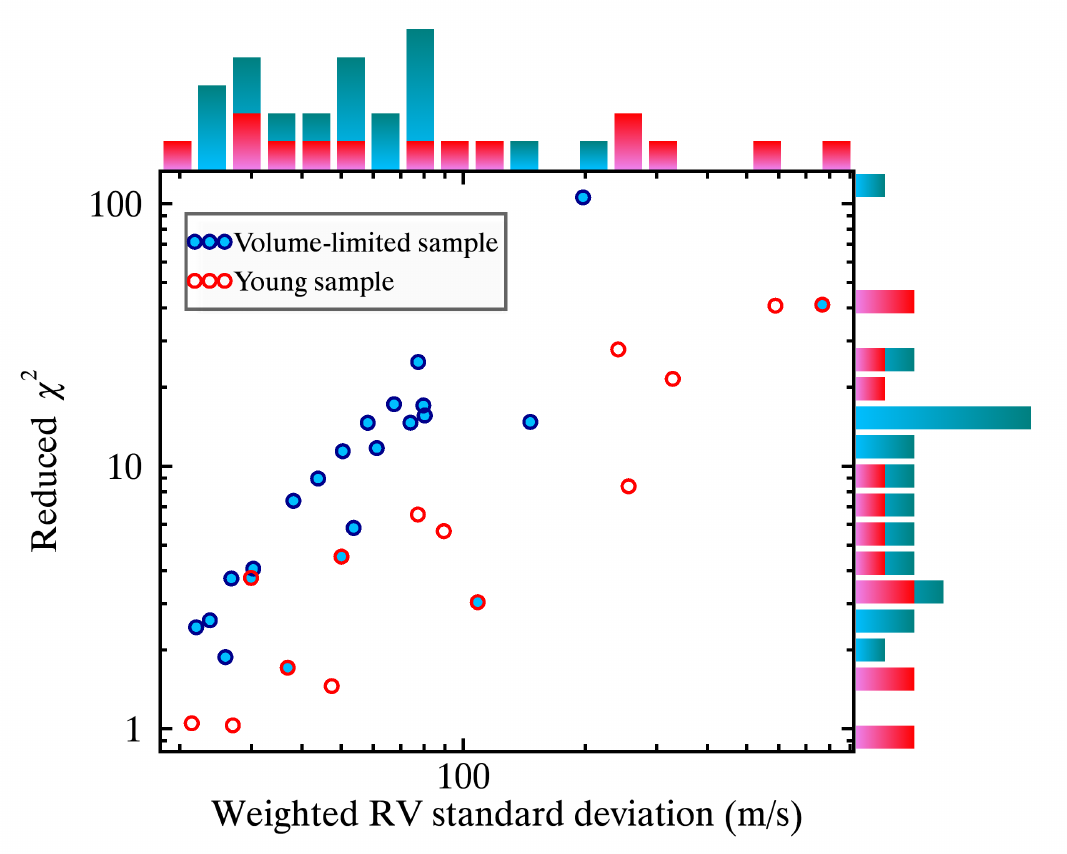} 
 \vspace*{-0.2 cm}
 \caption{Left: CSHELL NIR RV time-series for GJ 338A (N$_{obs}$=6, rms=22 m/s, $\chi_r^2=2.4$).  Right: CSHELL NIR RV survey results, plotted as a function of RV rms and reduced chi-squared for both young and nearby M dwarfs in red and blue respectively.}
\end{center}
\end{figure}

\begin{figure}
\vspace*{-0.1 cm}
\begin{center}
\includegraphics[width=2.0in,trim=6cm 2cm 2cm 2cm]{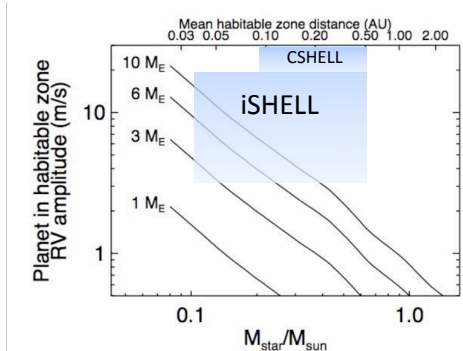} 
\vspace*{-0.2 cm}
 \caption{A comparison of the potential exoplanet sensitivity of CSHELL and iSHELL. The solid black lines correspond to the RV semi-amplitude of Habitable Zone (HZ) exoplanets with the indicated masses as a function of stellar mass.}
\end{center}
\end{figure}

\end{document}